\documentstyle[epsf]{mn}

\newif\ifAMStwofonts
\AMStwofontstrue

\ifoldfss
  \ifCUPmtlplainloaded \else
    \NewTextAlphabet{textbfit} {cmbxti10} {}
    \NewTextAlphabet{textbfss} {cmssbx10} {}
    \NewMathAlphabet{mathbfit} {cmbxti10} {} 
    \NewMathAlphabet{mathbfss} {cmssbx10} {} 
  \fi
  \ifAMStwofonts
    \ifCUPmtlplainloaded \else
      \NewSymbolFont{upmath} {eurm10}
      \NewSymbolFont{AMSa} {msam10}
      \NewMathSymbol{\upi}     {0}{upmath}{19}
      \NewMathSymbol{\umu}     {0}{upmath}{16}
      \NewMathSymbol{\upartial}{0}{upmath}{40}
      \NewMathSymbol{\leqslant}{3}{AMSa}{36}
      \NewMathSymbol{\geqslant}{3}{AMSa}{3E}

    \fi
  \fi
\fi 

\ifnfssone
  \newmathalphabet{\mathit}
  \addtoversion{normal}{\mathit}{cmr}{m}{it}
  \addtoversion{bold}{\mathit}{cmr}{bx}{it}
  \newmathalphabet{\mathbfit} 
  \addtoversion{normal}{\mathbfit}{cmr}{bx}{it}
  \addtoversion{bold}{\mathbfit}{cmr}{bx}{it}
  \newmathalphabet{\mathbfss} 
  \addtoversion{normal}{\mathbfss}{cmss}{bx}{n}
  \addtoversion{bold}{\mathbfss}{cmss}{bx}{n}
  \ifAMStwofonts
    \ifCUPmtlplainloaded \else
      \UseAMStwoboldmath
      \makeatletter
      \new@mathgroup\upmath@group
      \define@mathgroup\mv@normal\upmath@group{eur}{m}{n}
      \define@mathgroup\mv@bold\upmath@group{eur}{b}{n}
      \edef\UPM{\hexnumber\upmath@group}
      \new@mathgroup\amsa@group
      \define@mathgroup\mv@normal\amsa@group{msa}{m}{n}
      \define@mathgroup\mv@bold\amsa@group{msa}{m}{n}
      \edef\AMSa{\hexnumber\amsa@group}
      \makeatother
      \mathchardef\upi="0\UPM19
      \mathchardef\umu="0\UPM16
      \mathchardef\upartial="0\UPM40
      \mathchardef\leqslant="3\AMSa36
      \mathchardef\geqslant="3\AMSa3E
    \fi
  \fi
\fi 

\ifnfsstwo
  \DeclareMathAlphabet{\mathbfit}{OT1}{cmr}{bx}{it}
  \SetMathAlphabet\mathbfit{bold}{OT1}{cmr}{bx}{it}
  \DeclareMathAlphabet{\mathbfss}{OT1}{cmss}{bx}{n}
  \SetMathAlphabet\mathbfss{bold}{OT1}{cmss}{bx}{n}
  \ifAMStwofonts
    \ifCUPmtlplainloaded \else
      \DeclareSymbolFont{UPM}{U}{eur}{m}{n}
      \SetSymbolFont{UPM}{bold}{U}{eur}{b}{n}
      \DeclareSymbolFont{AMSa}{U}{msa}{m}{n}
      \DeclareMathSymbol{\upi}{0}{UPM}{"19}
      \DeclareMathSymbol{\umu}{0}{UPM}{"16}
      \DeclareMathSymbol{\upartial}{0}{UPM}{"40}
      \DeclareMathSymbol{\leqslant}{3}{AMSa}{"36}
      \DeclareMathSymbol{\geqslant}{3}{AMSa}{"3E}
    \fi
  \fi
\fi 

\ifCUPmtlplainloaded \else
  \ifAMStwofonts \else 
    \def\upi{\pi}
    \def\umu{\mu}
    \def\upartial{\partial}
  \fi
\fi

\makeatletter
\def\figsize{\ifSFB@referee\epsfxsize=0.5\hsize\else\epsfxsize=\hsize\fi}
\makeatother
\gdef\0{\phantom{0}}
\def\n(#1){{}$^{\rm(#1)}$}
\def\eq#1 {\begin{equation} #1 \end{equation}}

\def\Ref{\bibitem{}}
\def\symbol#1{\hbox{$#1$}}

\def\about {\symbol{\sim}}
\def\Tstar {\symbol{T_\ast}}
\def\Tin   {\symbol{T_1}}  
\def\Td    {\symbol{T_{\rm d}}}
\def\rin   {\symbol{r_1}}
\def\rout  {\symbol{r_2}}
\def\mic   {\symbol{\umu{\rm m}}}
\def\Fl    {\symbol{F_\lambda}}

\pagerange{\pageref{firstpage}--\pageref{lastpage}}
\pubyear{1996}

\title[Benchmark problems for dust radiative transfer]
 {Benchmark problems for dust radiative transfer}

\author[Ivezi\'{c}, Groenewegen, Men'shchikov and Szczerba]
{\v{Z}. Ivezi\'{c}$^1$, M.A.T. Groenewegen$^2$, A. Men'shchikov$^3$ and R. Szczerba$^4$ \\
  $^1$ Department of Astrophysical Sciences,
       Princeton University, Princeton, NJ 08544-1001, USA\\
  $^2$ Max-Planck-Institut f\"ur Astrophysik,
       Karl-Schwarzschild-Stra{\ss}e 1, D-85748 Garching, Germany\\
  $^3$ Polish Academy of Sciences, N. Copernicus Astronomical Center,
       00-716 Warsaw, Bartycka 18, Poland \\
  $^4$ Polish Academy of Sciences, N. Copernicus Astronomical Center,
       87-100 Toru\'n, Rabia\'nska 8, Poland}

\date{Accepted 1997. Received 1997; in original form 1997}

\begin{document}

\label{firstpage}

\maketitle

\begin                      {abstract} 
When verifying a sophisticated numerical code, it is a usual practice
to compare the results with reliable solutions obtained by other
means.  This work provides such solutions for the wavelength dependent
dust radiative transfer problem. We define a set of benchmark problems
in spherical geometry and solve them by three radiative transfer codes
which implement different numerical schemes. Results for the dust
temperature and emerging spectra agree to better than 0.1\%, and can be
used as benchmark solutions for the verification of the dust radiative
transfer codes.
\end{abstract}


\section                    { INTRODUCTION }

Dust is abundant in the universe and many astronomical objects are
associated with it. Most notable examples are galaxies and stars, at
almost all points in their evolution. Dust surrounding an embedded
source scatters, absorbs, and reemits radiation originating from the
source. This processing usually results in an overall shift of spectral
energy distribution to long, infrared (IR) wavelengths\footnote{This is
strictly true only in spherical geometry.  Sources with toroidal, or
disk-like dust distribution can show significant amounts of short
wavelength radiation if viewed face-on.}.  Embedded sources may be
entirely obscured by the dust at optical wavelengths, and the only
available information is obtained at IR wavelengths, invisible from
earth until recently.

Thanks to the recent developments in IR techniques and facilities, the
spectral energy distribution for many objects is now available, and the
amount and quality of data are steadily increasing. Interpretation of
these data can offer insight to the nature of optically obscured
objects, making their IR signature a powerful tool of analysis.
However, because of the complexity of the radiative transfer problem
such analysis must be aided by sophisticated computational tools.

The wavelength dependent dust radiative transfer problem can be solved
only numerically and early attempts to obtain approximate analytical
and semi-analytical solutions assumed gray opacity (e.g.  Chandrasekhar
1934, Kozirev 1934). These early works were based on the Eddington
approximation that the radiation field is isotropic. The first direct
numerical solutions were obtained by Hummer \& Rybicki (1971) who
employed iterations over variable Eddington factor (the ratio of the
second to the zeroth moment of the radiation intensity). However, they
also assumed a gray opacity, an approximation too crude for detailed
analysis of observations.

The first calculations of the wavelength dependent dust radiative
transfer in spherical geometry were performed by Scoville \& Kwan
(1976), and Leung (1976). Both attempts included some unrealistic
assumptions and the first full approximation-free solution was obtained
by Rowan-Robinson (1980, hereafter RR). His method is based on the
direct integration of the radiative transfer equation, also known as
ray tracing. Yorke (1980) developed a method based on iterations over
variable wavelength dependent Eddington factors, a non-gray extension
of the method originally used by Hummer and Rybicki. Both approaches
remained the most widely used methods in subsequent developments of new
codes (e.g., Groenewegen 1993) whose number has been steadily
increasing during last two decades. Gradually, the codes' capabilities
have been extended to two-dimensional geometries (e.g., Efstathiou \&
Rowan-Robinson 1990, Collison \& Fix 1991, Men'shchikov \& Henning
1997), and even to the three-dimensional case (Steinacker \& Henning
1996).  Yet, an important shortcoming remains in this development.

When verifying a sophisticated numerical code, it is a usual practice
to compare the results with an analytical solution, or with a reliable
solution obtained by other means. There is no such solution available
for the wavelength dependent continuum radiative transfer problem.
Definition of such a radiative transfer problem involves several
complicated functions as its input (e.g., the spectrum of the embedded
source, wavelength dependent grain optical properties), making
analytical solutions impossible. The only practical approach is to
compare solutions obtained by several independent codes for a set of
well defined problems. If all such codes produce the same results, then
it is likely that these results are reliable, and can be used as
solutions of the benchmark problems for code verification.

This work defines a set of such benchmark problems in spherical
geometry and compares solutions obtained by three radiative transfer
codes which implement different numerical schemes. The
benchmark problems are defined in Section 2, where we also briefly
describe the radiative transfer codes which we used. Solutions for the
dust temperature and emerging spectra are presented in Section 3, as
well as the discussion of our results.

\section{BENCHMARK PROBLEMS}

Most of works describing infrared emission from astronomical sources
present various computational results. Yet, it is difficult to use
those results for the code verification. While many of the model
parameters are usually explicitly listed, very often it is not easy to
recognize all the underlying assumptions, nor to reproduce the grain
optical properties used in the calculations. To avoid these problems,
all input properties for the benchmark problems presented here are
defined analytically. Such approach will ease comparison of the results
obtained by other codes with those presented here.

\subsection{Scaling Properties of the Radiative Transfer Problem}

Traditionally, detailed modeling of IR radiation involved numerous 
input quantities, necessitating a large number of calculations to 
obtain successful fits, and diluting the value of the resulting 
success. However, as pointed out by RR, the number of relevant 
parameters can be drastically reduced by employing the scaling 
properties of the radiative transfer problem. For example, the 
luminosity of the central source is irrelevant, a fact by and large 
ignored in modeling astronomical sources. 

The importance of scaling was recently emphasized by Ivezi\' c \&
Elitzur (1995) who demonstrated that for given dust type, the IR
emission from late-type stars can be successfully described by a single
parameter --- the overall optical depth $\tau$\footnote{Assuming a
steady-state radiatively driven wind.}.  All other quantities
(luminosity, mass, mass-loss rate, etc.) enter only indirectly through
their effect in determining $\tau$, and thus are irrelevant in the
modeling of IR emission.  It was subsequently recognized that this
powerful scaling is a general property of radiative transfer and that
it can be extended to arbitrary geometries (Ivezi\' c \& Elitzur 1997,
hereafter IE97).  IE97 point out that the spectral shape is the only
relevant property of the heating radiation when the inner boundary of
the dusty region is controlled by dust sublimation.  Similarly, the
absolute scales of densities and distances are irrelevant; the geometry
enters only through angles, relative thicknesses and aspect ratios. The
actual magnitudes of densities and distances enter only through one
independent parameter, the overall optical depth.  Dust properties
enter only through dimensionless, normalized distributions that
describe the spatial variation of density and the wavelength dependence
of scattering and absorption efficiencies. We now proceed to define the
benchmark problems in terms of these fully scaled quantities.

\subsection{Definition of the Benchmark Problems}

A central point source is embedded in a spherically symmetric
dusty envelope with an inner cavity free of dust. The source radiates as 
a black body at a given temperature \Tstar. The dust is in radiative 
equilibrium with the local radiation field, and this uniquely determines the 
dust temperature, \Td, at every point in the envelope. The scale of \Td\ 
is determined by \Tin, the dust temperature at the inner boundary, \rin.
All radial positions can be scaled by this radius defining a new,
dimensionless variable $y=r/\rin$. The dimensionless outer radius of the 
envelope, $Y = \rout/\rin$, is a free parameter. The dust density variation
with $y$ is assumed  to be a power law $\propto y^{-p}$. The actual dust 
density, envelope size and opacity scales are all combined into a value for 
overall optical depth specified at some fiducial wavelength. We choose 1 
\mic\ for this wavelength and denote the corresponding optical depth by 
$\tau_1$.

Dust optical properties are specified as scale-free wavelength
dependent absorption and scattering opacities normalized to unity at 1
\mic, $q_{\rm abs}=\kappa_{\lambda,{\rm abs}}/\kappa_{1,{\rm abs}}$ and
$q_{\rm sca}=\kappa_{\lambda,{\rm sca}}/\kappa_{1,{\rm sca}}$,
respectively.  For spherical amorphous grains with radius $a$, $q_{\rm abs}$ 
and $q_{\rm sca}$ are roughly constant for $\lambda < 2\pi a$, and fall off
as $\lambda^{-1}$ and $\lambda^{-4}$ for $\lambda > 2\pi a$,
respectively. To mimic this behavior\footnote{We assume isotropic
scattering.} with analytical functions, we choose
\begin{equation}
  q_{\rm abs} =  q_{\rm sca} = 1      
\end{equation}
for $\lambda < 1 \mic$, and
\begin{equation}
  q_{\rm abs} = {1 \over \lambda  }  
\end{equation}    
\begin{equation}          
  q_{\rm sca} = {1 \over \lambda^4 }     
\end{equation}
for $\lambda > 1 \mic$. The arbitrary choice of 1 \mic\ as transitional 
wavelength is motivated both by typical astronomical grain sizes, and by desire
to have such a transition at short wavelengths where its effects are most
easily discernible. The chosen forms correspond to the maximal theoretically 
possible efficiencies for spherical amorphous grains with radius 0.16 \mic\ 
(Greenberg 1971).  

The above listed quantities fully specify the benchmark problems.
We perform calculations for \Tstar = 2500 K, \Tin = 800 K and $Y$ = 1000,
and produce eight different models: for two different density distributions,
$p=2$ and $p=0$, and four values of optical depth, $\tau_1$ = 1, 10, 100, 1000.
Selection of different density distributions and a range in optical depth
minimizes the chance that eventual inconsistencies in different numerical
schemes would not be noticed.

\subsection{Code Description}

The three codes used in this work have similar approach to the solution
of the radiative transfer problem, and none of them introduces any
approximations (e.g., closure relation in the moment methods, Auer
1984). Since the problem cannot be solved analytically, the solutions
are obtained on discrete spatial and wavelength grids. Within arbitrary
numerical accuracy which determines the grid sizes, the solutions can
be considered exact.

The spatial grids include the radial position and impact parameter grids,
which also directly determine the angular grid needed for the integrations
over solid angle. Sizes of the spatial grids range from \about 10 to several
hundred points, depending on the method and overall optical depth for a given
model. The wavelength grid has typically \about 100 points. To solve
the radiative transfer problem means to determine the radiation intensity
at each of the grid points, that is, to determine the wavelength 
dependent intensity as a function of angle for every radial position.
 
The radiative transfer equation is an integro-differential equation; in
other words, the intensity at any grid point depends on the intensities at
all other grid points. Thus the problem cannot be solved by straightforward 
techniques, and various numerical methods rely on iterations of 
different types. These iterations can be performed either for the intensity,
or for its moments (energy density, flux, pressure, etc.). The codes
used in this work implement different numerical schemes to perform 
the iterations, and we proceed with their brief descriptions.

\bigskip 
{\bf Code 1} This code was originally developed by Yorke (1980) and
generalized by Men'shchikov \& Henning (1997) and Szczerba et al.
(1996, 1997). The original version solves differential moment equations
obtained by analytically integrating the radiative transfer equation
multiplied by the powers of the direction cosine, over the solid angle
(e.g. Auer 1984). Such an approach always results in a number of
differential equations of one less than the number of unknown
quantites (the moments of the radiation intensity).  The closure
relation can be expressed in terms of the variable Eddington factor,
and this code calculates it directly from its defining equation after
every iteration. Later versions of the code were extended to improve
explicit ray tracing, with iterations repeated until convergence in
dust temperature and mean intensity is achieved.

\bigskip
{\bf Code 2} Groenewegen (1993) has developed a code which solves the
radiative transfer equation in spherical geometry from first
principles, assuming isotropic scattering. The intensity is evaluated
explicitly on sufficiently fine grids to obtain desired accuracy, and
iterations are repeated until the convergence is achieved.  This code
was developed to allow for an explicit mass-loss rate dependence on
time and velocity law, rather than a power-law density distribution.
The model was tested against the results obtained with the model of
Rogers \& Martin (1984, 1986), for silicate grains up to an optical
depth at 9.5 $\mu$m of 50; differences were 1\% at most.  For the
purpose of the present paper, models with $p$ = 2 (constant expansion
velocity and mass loss rate) and ${\tau}_1$ = 1, 10, 100 were
calculated (${\tau}_1$ = 1000 was not included for this 
code because of unsatisfactory convergence).

\bigskip
{\bf Code 3} The third code, DUSTY, was developed by Ivezi\' c, Nenkova
\& Elitzur (1997) and is publicly available. It solves the integral
equation for the energy density obtained by analytically integrating
the radiative transfer equation. The subsequent numerical integration
is transformed into multiplication with a matrix of weight factors
determined purely by the geometry. The energy density at every point is
then determined by matrix inversion, obviating the need to iterate over
the energy density itself. That is, unlike other codes, DUSTY can
directly solve the pure scattering problem. The intensity is not
explicitly evaluated, and can be easily recovered from the source
function by using the weight factor matrix.

\begin{table}
\begin{center}

\caption{Dimensionless parameter $\Psi$ for eight benchmark models}

\smallskip
\begin{tabular}{@{}ccccc}
\hline
   p\n(a) & $ \tau_1 = 1 $ & $\tau_1 = 10 $ & $\tau_1 = 100 $ & $\tau_1 = 1000 $ \\
\noalign{\smallskip}
   2  &   3.48    &  5.42    &    13.1  &   84.1    \\
   0  &   2.99    &  3.00    &    3.10  &   3.75    \\   \hline
\end{tabular}
\end{center}
\smallskip

(a) Power for the power-law describing the dust density distribution

\end{table}

\begin{figure}
\centering \leavevmode \figsize \epsfbox[70 80 570 780]{Temp.ps}
\caption{Dust temperature distribution through the envelope for two
density distributions and optical depths at 1 \mic\ as marked.}

\end{figure}

\section{RESULTS}

The full solution of the radiative transfer problem is contained in the
radiation intensity. However, even in the spherically symmetric systems
it depends on three variables (position, angle, and wavelength) and its
presentation would be quite involved. When isotropic scattering
is assumed as here, the solution is also fully described by the energy
density since it fully defines the source function. Equivalently,
the emerging spectrum, the dust temperature distribution, and 
dimensionless parameter 
\eq{}{
             \Psi = {4\sigma \Tin^4 \over F_1},
}
where $\sigma$ is the \v Stefan-Boltzman constant and $F_1$ is the 
bolometric flux at $y$=1, also fully specify the solution\footnote{The
absolute size of the inner boundary can be determined using $\Psi$,
see eq. (27) of IE97} (IE97). That is, if these three quantities
obtained by different codes agree, then the intensity distributions 
at every grid point agree, too.

\begin{figure}
\centering \leavevmode \figsize \epsfbox[70 80 570 780] {Spec.ps}
\caption{Spectral energy distribution of the emerging radiation
for two density distributions and optical depths at 1 \mic\ as marked.}

\end{figure}

The values for $\Psi$ obtained for eight models calculated in this
work are given in Table 1. All three codes agree within significant
digits listed. Temperature distributions are presented in Figure 1.
Again, the agreement is better that 0.1\%. Emerging spectra are
shown in Figure 2. They are presented as dimensionless, distance 
and luminosity independent spectral shape 
$\lambda \Fl / \int{\Fl d\lambda}$. Differences between the results 
obtained by different codes are smaller than the line thickness 
($<$ 0.1\%). 

The detailed numerical values for emerging spectra presented in  
Figures 1 and 2, and for all other relevant quantities can be 
obtained in computer readable form from \v Z. Ivezi\' c\footnote{E-mail
address: ivezic@astro.princeton.edu.}. These 
results can be used for the verification of the wavelength dependent 
radiative transfer codes, especially for the new multi-dimensional
ones. Also, as the numerical radiative-hydrodynamical codes for 
modeling star and planet formation begin to include the wavelength 
dependent radiation transfer, the benchmark problems presented here 
might prove valuable for establishing confidence in the accuracy of 
their results (Boss 1996).

\section*{Acknowledgments}

We thank M. Elitzur and G. Knapp for their careful reading, A. Boss for his 
encouragement, and the referee C. Skinner for useful comments which helped 
improve the manuscript.

\label{lastpage}

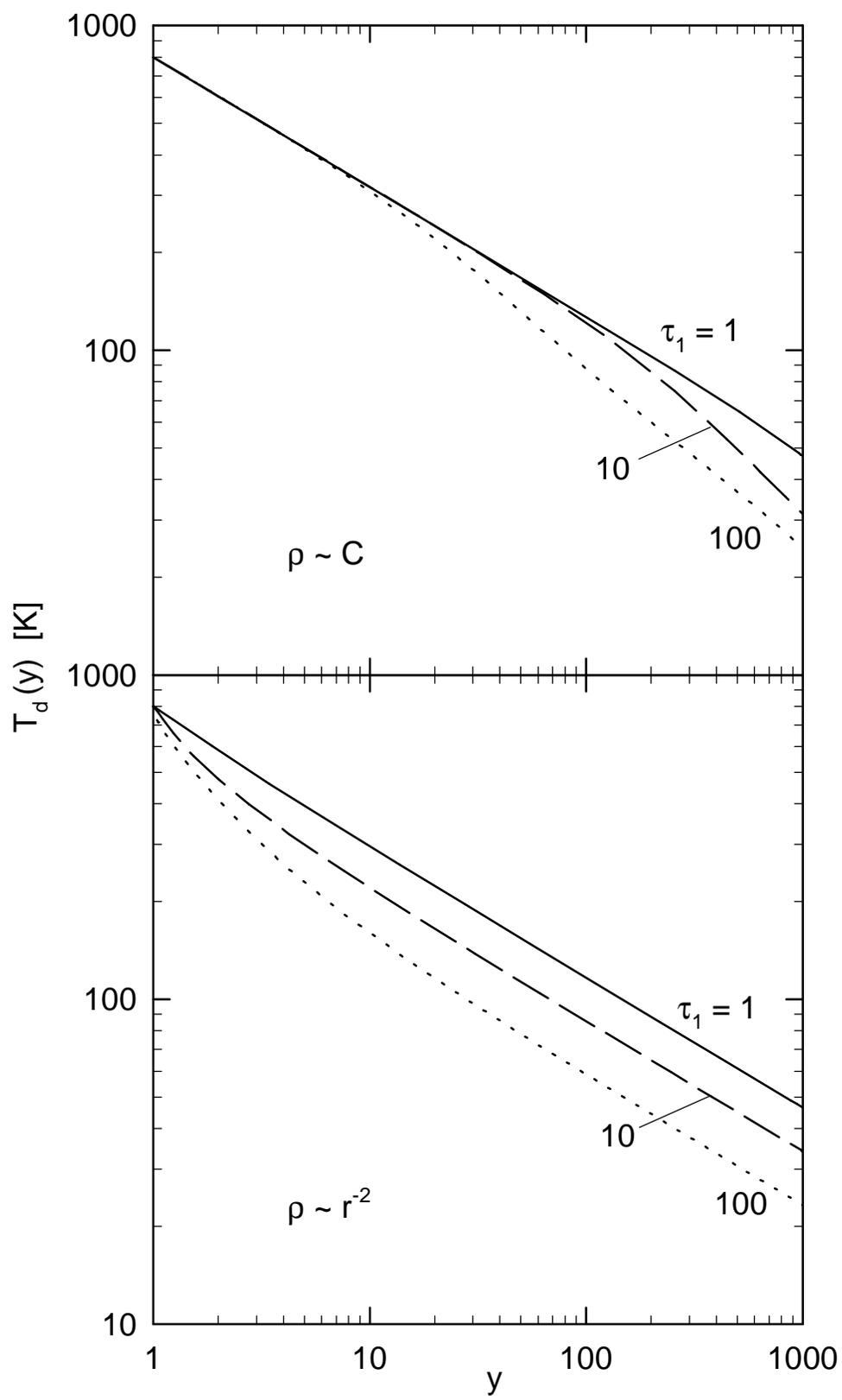
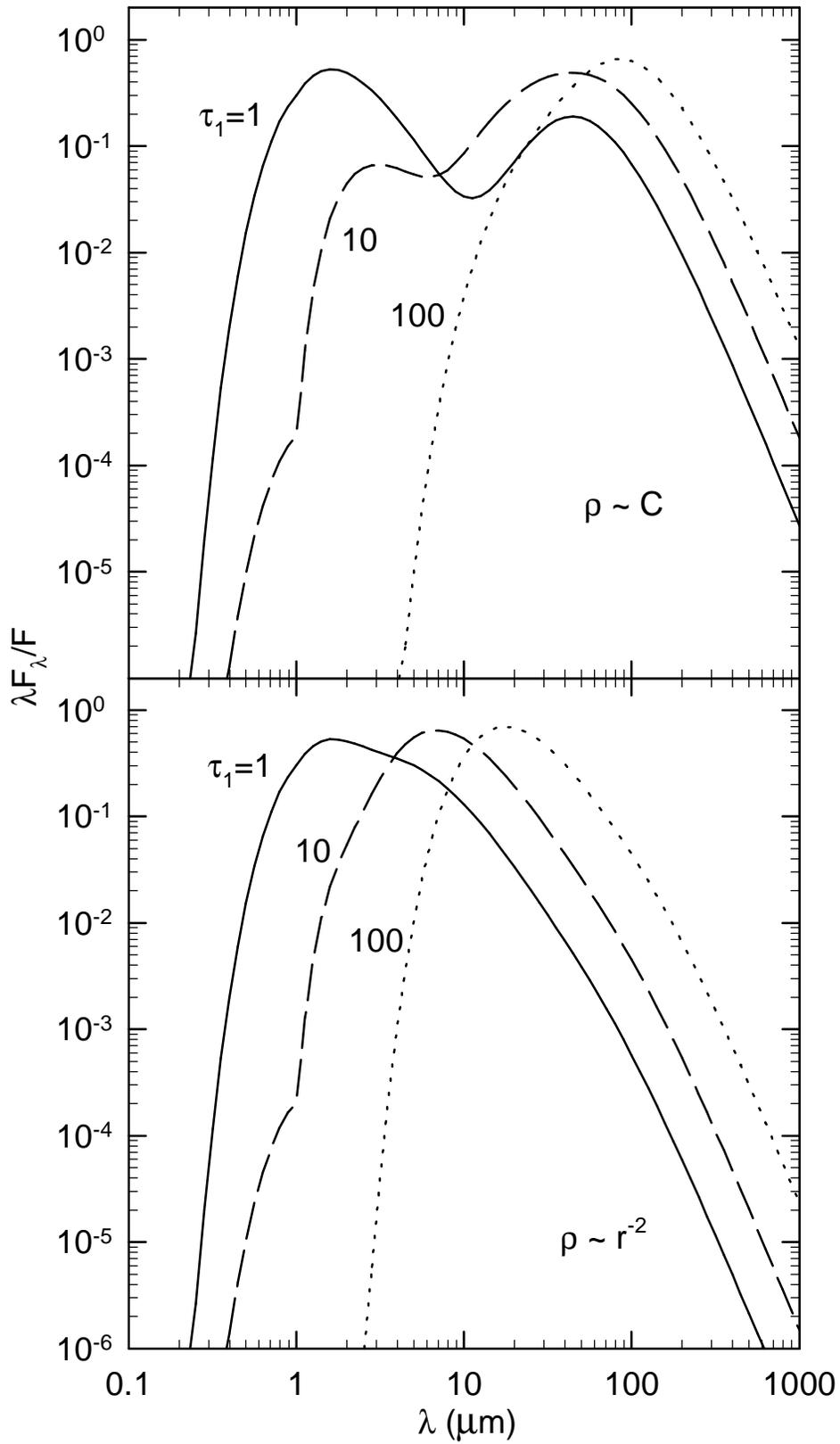
\end{document}